\documentclass[pdftex,twocolumn,epjc3]{svjour3}          

\RequirePackage[T1]{fontenc}

\smartqed  

\RequirePackage{graphicx}
\RequirePackage{mathptmx}      
\RequirePackage{flushend}
\RequirePackage[numbers,sort&compress]{natbib}
\RequirePackage[colorlinks,citecolor=blue,urlcolor=blue,linkcolor=blue]{hyperref}

\journalname{TTMP}

	[2011/08/26 v1.2
	style option for the EPJC journal]
	
	\AtEndOfClass{\@input{svepjc3.clo}}
	\gdef\@epjcloaded{}

	\AtBeginDocument{%
	\def\@xhline{\ifx\reserved@a\hline
		\vskip\doublerulesep\vskip-\arrayrulewidth\fi
		\ifnum0=`{\fi}%
		\noalign{\vskip2\p@}}%
}

\begin{document}
	
	\title{Solutions of generalized supergravity equations with the BTZ black hole metric }

	\author{Ali Eghbali\thanksref{e1}
		\and
        Simin Ghasemi-Sorkhabi\thanksref{e2}
        \and
		Adel Rezaei-Aghdam\thanksref{e3} 
	}
	
	\thankstext{e1}{Corresponding author: eghbali978@gmail.com}
	\thankstext{e2}{s.ghassemi.s@gmail.com}
    \thankstext{e3}{rezaei-a@azaruniv.ac.ir}
	
	\institute{Department of Physics, Azarbaijan Shahid Madani University, 53714-161, Tabriz, Iran}
	
	\date{Received: date / Accepted: date}

	\maketitle

\section*{Abstract}
We proceed to investigate the solutions of generalized supergravity equations (GSE) in three dimensions.
Our candidate is the metric of BTZ black hole. It is shown that only the cases with $J=M=0$ and $J=0,~ M\neq 0$ of the BTZ metric satisfy the GSE.
In the former, we find a family of solutions including the field strength $H_{_{r \varphi t}}=2r/l$,
the cosmological constant $\Lambda=-1/l^2$,  one-form $Z_{\mu}$ and a vector field which is obtained to be a linear combination of the directions of the time translation and rotational symmetries. In the latter, the solutions possess the same field strength as before, while the cosmological constant $\Lambda$,
one-form $Z_{\mu}$ and vector field $I$ will be different from the previous case.
Finally, we show that the charged black string solution
found by Horne and Horowitz, which is Abelian T-dual to the the BTZ black hole solution, can be considered as a solution for the GSE.
\\\\
{\bf Keywords:}  Generalized supergravity equations; Standard supergravity; BTZ metric
\section{\label{I} Introduction}
Supergravity is a modern field theory combining the principles of supersymmetry and general relativity. The 10-
dimensional supergravity theory describes the dynamics of massless string excitations and arises in string theory as a low-energy effective theory.
Recently, new string backgrounds have been discovered that satisfy a more general set of motion equations than ordinary supergravity.
These equations, which are a generalization of the standard supergravity equations, are called the GSE.
It's worth noting that the primary difference between standard supergravity and GSE is the absence of a scalar dilaton.
Arutyunov, {\it et al.} \cite{r1} introduced the GSE in order to investigate integrable deformations of
the $AdS_{5}\times S^{5}$ type II superstring sigma model \cite{r2,r3}\footnote{In order to construct the Yang-Baxter deformations of $AdS_5 \times S^5$, one must employ the classical r-matrix as a solution of the homogeneous classical Yang-Baxter equation \cite{r4}.
It has been shown that the Yang-Baxter deformed background of $AdS_5 \times S^5$ satisfies the equations of motion of type IIB supergravity if
the classical r-matrix satisfies the unimodularity condition \cite{r5} (see, also, \cite{{Sheikh-Jabbari1},{Sheikh-Jabbari2},{Sheikh-Jabbari3},{Sheikh-Jabbari4}}). Otherwise, the background is a solution of the GSE.}, which is closely related to non-Abelian T-duality transformations \cite{r6,r7,r8,r9}. This generalized system in string theory includes additional vector fields $I$ in addition to the standard component fields of type IIB supergravity. Up to now, the corresponding classical action has not been discovered, and only the equations of motion are presented.
Additionally, Ref. \cite{r11} presents a solution of standard supergravity with a linear dilaton that has been mapped to a GSE solution through a formal T-duality transformation along a specific direction. These findings emphasize the importance of treating solutions of standard supergravity and GSE equally in the context of string theory, as T-duality is a fundamental symmetry in string theory.
As a great progress in the recent study of string theory, Tseytlin and Wulff \cite{r10} demonstrated that the GSE can be reproduced by solving the $\kappa$-symmetry constraints. In fact, their results show that the $\kappa$-symmetry of the Green-Schwarz action requires the background supergravity fields to satisfy the GSE.

Later, the GSE attracted the attention of many researchers, in such a way that it was shown that \cite{yuho1} the
whole bosonic part of the type II GSE can be reproduced from the T-duality covariant equations of motion of the double field theory when one chooses
a non-standard solution of the strong constraint. Additionally, in Ref. \cite{yuho1},
the Weyl invariance of the bosonic sigma model on a generalized gravity background
has been shown by using the doubled formalism. We think that the results of work of Ref. \cite{yuho1} have provided
positive evidence that superstring theories defined on solutions of the GSE are Weyl invariant (see, also, \cite{yuho1.1}).
In Ref. \cite{yuho2}, without introducing a T-duality manifest formulation of string theory, it has been
constructed a local counterterm that cancels out the Weyl anomaly of bosonic string
theory defined in generalized supergravity backgrounds; the results of their work support the Weyl invariance of string theory in generalized supergravity backgrounds.
Also, it has been shown that the equations of motion of generalized supergravity can be followed from
the sigma model once the Killing vector $I$ is identified with the trace of the structure constants \cite{r8}.
In this regard, for the Bianchi spacetimes, it was shown that \cite{r8}
the non-Abelian T-duals with respect to non-semisimple groups are solutions to generalized supergravity.

As a spin off from this progress, it would be interesting to present new solutions for the GSE in three dimensions.
In this work, we obtain some new solutions for the GSE, including special cases of the BTZ metric ($J=M=0$ and $J=0,~ M\neq 0$).
Notably, when we select the vector field $I$ based on our perspective, it becomes evident that the cases with
$J\neq 0$,~$M=0$ and $J\neq 0,~M\neq0$ do not satisfy the GSE. As an interesting result, we show that the charged black string solution
found by Horne and Horowitz \cite{Horowitz1}, which is Abelian T-dual to the the BTZ black hole solution, can be considered as a solution for the GSE.

The structure of the paper is as follows. We start with a short overview of the GSE in section \ref{II}, and introduce
our notation. Next, in section \ref{III}, after the introduction of the BTZ black hole metric,
we consider the BTZ black hole solutions in the context of the low energy approximation.
Section \ref{IV} contains the original results of the work:
we look into the BTZ metrics in the context of the GSE and show that only the cases $J=M=0$ and  $J=0,~ M\neq 0$ of the BTZ metric can satisfy
the GSE.
In section \ref{V}, the charged black string solution is considered as a solution for the GSE.
Finally, conclusion is reported in section \ref{VI}.	

\section{\label{II} A brief review of the GSE}
The subject of this section is a brief overview of the GSE.
In the absence of  Ramond-Ramond fields, these equations in D dimensions take the following form \cite{r1}
\begin{eqnarray}
R_{\mu \nu}-\frac{1}{4}~H_{\mu \rho \sigma}~H^{\rho \sigma}_{\nu }+(\nabla _{\mu}~X_{\nu}+\nabla_{\nu}~X_{\mu})=0,\label{eq1}\\
\frac{1}{2}~\nabla^{\lambda}~H_{\lambda \mu \nu}-X^{\lambda}~H_{\lambda \mu \nu}-\nabla _{\mu}~X_{\nu}+\nabla_{\nu}~X_{\mu}=0,\label{eq2}\\
 R-\frac{1}{12}H^2+4\nabla_{\mu}~X^{\mu}-4X_{\mu}~X^{\mu}-4~\Lambda=0,\label{eq3}
\end{eqnarray}
where $R_{\mu \nu}$ and $R$ are the respective Ricci tensor and Gauss curvature that are calculated
from the metric $G_{\mu \nu}$, and $\Lambda$ is the cosmological constant.
Here, the $D$-dimensional indices $\mu,~\nu,...$ of coordinates $x^{\mu}$ are raised or lowered with the metric $G_{\mu \nu}$.
The covariant derivative $\nabla _{\mu}$ is the conventional Levi-Civita connection associated with $G_{\mu \nu}$.
The field strength $H_{\mu \nu \rho}$ corresponding to anti-symmetry tensor field B is defined as
\begin{equation}
H_{\mu \nu \rho}=\partial_{\mu}~B_{\nu \rho}+\partial_{\nu}~B_{\rho \mu}+\partial_{\rho}~B_{\mu \nu}.\label{eq4}
\end{equation}
In addition, $X_{\mu}$ is defined to be $X_{\mu}=I_{\mu}+Z_{\mu}$ in which $I=I^{\mu}~\partial_{\mu}$ is
a vector field, while $Z=Z_{\mu}~dx^{\mu}$ is a one-form. They satisfy \cite{r1}
\begin{eqnarray}\label{eq4}
       \mathcal{L}_{_I}~G_{\mu \nu}=0,\label{eq5}\\
       \mathcal{L}_{_I}~B_{\mu \nu}=0,\label{eq6}\\
       \nabla _{\mu}~Z_{\nu}-\nabla_{\nu}~Z_{\mu}+I^{\lambda}~H_{\lambda \mu \nu}=0,\label{eq7}\\
       I^{\lambda}~Z_{\lambda}=0,\label{eq8}
\end{eqnarray}
where $\mathcal{L}$ stands for the Lie derivative.
The conventional dilaton is included in $Z_{\mu}$ as follows:
\begin{equation}
Z_{\mu }=\partial_{\mu}~\Phi+B_{\nu \mu}~I^{\nu}.\label{eq9}
\end{equation}
Here, $\Phi$ is a scalar dilaton field hidden within $Z_{\mu}$.
Note that if we set $I^{\mu}=0$, then one gets that  $X_{\mu} = \partial_{\mu}~\Phi$, and thus, the GSE reduce to the standard supergravity equations.

In this work, we present new solutions for the GSE (\ref{eq1})-(\ref{eq3}) together with (\ref{eq5})-(\ref{eq8}),
including special cases of the BTZ metric, field strength $H$, one-form $Z_{\mu}$, and an appropriate vector field $I$.
In this manner, we will derive a family of solutions for the cases $J=0,~M=0$ and $J=0,~ M\neq 0$ within the BTZ metric.
As mentioned earlier, the case $I^{\mu}=0$ of the GSE corresponds to the standard supergravity equations.
It has already been shown that \cite{Horowitz} a slight modification of the BTZ black hole solution
yields an exact solution to the standard supergravity equations. In the next section, after a brief overview of the BTZ black hole,
we consider the BTZ metric in the context of the low energy approximation.

\section{\label{III} The BTZ black hole metric as a solution of the standard supergravity equations}

First of all, let us introduce the metric of BTZ black hole. The BTZ black hole, discovered by Banados, Teitelboim, and Zanelli \cite{r12}, is a $2+1$-dimensional solution of Einstein's equations with negative cosmological constant, mass, angular momentum, and charge. Unlike its higher-dimensional counterparts, the BTZ black hole is asymptotically anti-de Sitter and lacks a curvature singularity at the origin.
The line element for the black hole solutions is as follows \cite{r12}:
\begin{eqnarray}
ds^2&=&(M-\frac{r^2}{l^2})~dt^2-J~dt~d\varphi+r^2~d\varphi ^2\nonumber\\
&+&(\frac{r^2}{l^2}-M+\frac{J^2}{4r^2})^{-1}~dr^2,~~~~~~0\leq \varphi \leq 2\pi.\label{eq10}
\end{eqnarray}
where the radius $l$ is related to the cosmological constant by $l = (-\Lambda)^{-1/2}$.
The constants of motion, denoted by $M$ and $J$, represent the mass and angular momentum of the BTZ black hole, respectively.
The line element (\ref{eq10}) describes a black hole solution with outer and inner horizons at $r=r_{+}$ and $r=r_{-}$ , respectively,
\begin{equation}\label{eq11}
r_{\pm }= l \bigl ({{M}\over 2}\bigr )^{1/2}
\biggl \{1\pm \biggl (1- {{J^2} \over {{M^2} l^2}} \biggr)^{1/2}\biggr \}^{1/2}.
\end{equation}
The mass $M$ and angular momentum $J$ are related to $r = r_{\pm}$ by $M = (r_{+}^{2} + r_{-}^{2})/l^2$ and $J = 2~r_{+}~r_{-}/l^2$. Solutions with $-1 < M < 0$ and $J = 0$ describe point particle sources with naked conical singularities at $r = 0$. The metric with $J = 0$ and $M = -1$ can be recognized as ordinary anti-de Sitter space, and it is separated by a mass gap from the case where $J = 0$ and $M = 0$. The vacuum state, representing empty space, is obtained by letting the horizon size go to zero. This corresponds to $M\rightarrow 0$, which requires $J\rightarrow 0$. It's worth noting that the metric for the $J = 0$ and $M = 0$ black hole is not the same as the $AdS_3$ metric, which has negative mass, $M = -1$.

As mentioned in the previous section, a modified BTZ black hole solution was
obtained to a $2 + 1$-dimensional string theory with a matter source given by anti-symmetric $B$-field with the contribution of the cosmological constant $\Lambda$ \cite{Horowitz}. There, Horowitz and Welch showed that very solution to 3-dimensional general relativity with
$\Lambda<0$ can be considered as a solution to the standard supergravity equations with a constant dilaton field, $\Lambda=-1/l^2$ and
field strength $H_{_{\mu \nu \rho}} =2 \epsilon_{_{\mu \nu \rho}}/l$, where $\epsilon_{_{\mu \nu \rho}}$ stands for the volume
form in three dimensions\footnote{Note that a special property of three dimensions is that the field strength
$H_{_{\mu \nu \rho}}$ must be proportional to the volume form $\epsilon_{_{\mu \nu \rho}}$.}.
In addition, it was shown that \cite{Horowitz} the solution of BTZ black hole (\ref{eq10}) along with
a constant dilaton field, $\Lambda=-1/l^2$ and field strength $H_{_{r \varphi t}} =2 r/l$
satisfy the equations of motion of the standard supergravity. It is worth mentioning that
3-dimensional black hole solutions to (\ref{eq10}) do not exist assumed that $H_{_{\mu \nu \rho}}=0$.

\section{\label{IV} Solutions of the GSE with the BTZ metric}

In what follows we shall look into the BTZ solutions in the context of the GSE.
We show that the cases $J=M=0$ and  $J=0,~ M\neq 0$ of the BTZ metric can be
considered as solutions of the GSE. As mentioned in section \ref{III}, the ${M}$ and $J$ are the mass and angular momentum of the BTZ black hole, respectively. They are appeared due to the time translation symmetry and rotational symmetry of the metric,
corresponding to the Killing vectors ${\partial / \partial t}$ and ${\partial/\partial \varphi}$, respectively.
On the other hand, relation (\ref{eq5}) is called Killing equation, which in terms of a Killing vector field $K_a=K_a^{~\mu} \partial_\mu$
it can be written as
\begin{eqnarray}\label{eq12}
\partial_{\mu} K_a^{~\lambda}  ~ G_{_{\lambda \nu}} +K_a^{~\lambda} \partial_{\lambda} G_{_{\mu \nu}}+\partial_{\nu} K_a^{~\lambda} ~ G_{_{\mu \lambda}}=0.
\end{eqnarray}
Accordingly, the vector field $I$ can be a Killing vector or a linear combination of the Killing
vectors corresponding to metric (\ref{eq10}). The Killing vectors $K_a$ corresponding to the BTZ metric can be
derived by solving Killing equation (\ref{eq12}), giving us six linearly independent vectors.
Here, to construct an appropriate vector field $I$ we use the Killing vectors of the BTZ metric. We assume that the choice of
\begin{eqnarray}\label{eq13}
I=\alpha_1 K_1 +\cdots + \alpha_6 K_6,
\end{eqnarray}
for some constants $\alpha_i$, can be a suitable candidate for solving the GSE (\ref{eq1})-(\ref{eq3}) together with (\ref{eq5})-(\ref{eq8}).
It should be noted the fact that the BTZ solutions must be single-valued in the angular direction.
Therefore, one should be careful in choosing the vector field $I$.
As the first example, we look into the case $J=0, M=-\lambda^2 <0$ of (\ref{eq10}) in full detail.

\subsection{Solutions with J=0, M=-$\lambda^2 <0$ }
In this subsection we shall investigate the solutions of the GSE for the case $J=0, M=-\lambda^2 <0$ of the BTZ metric.
Before proceeding to do this, let us first write down the BTZ metric for the case $J=0, M=-\lambda^2 <0$. Using relation (\ref{eq10}), it is given by
\begin{equation}\label{eq14}
ds^2=-(\lambda^2+\frac{r^2}{l^2})~dt^2+(\lambda^2+\frac{r^2}{l^2})^{-1}~dr^2+r^2~d\varphi^{2}.
\end{equation}
The Killing vectors of the metric (\ref{eq14}) can be derived by solving
Killing equations. They are then read off
\begin{eqnarray}
K_{1}  &=& \frac{\partial}{\partial {\varphi}},\nonumber\\
K_{2}  &=& \sqrt{\lambda^2 {l^2}+{r^2}}~~\Big[\frac{r l}{\lambda^2 {l^2} +r^2}~\cos(\frac{\lambda t}{l}) \sin(\lambda \varphi)~\frac{\partial}{\partial {t}} \nonumber\\
&& +\lambda \sin(\frac{\lambda t}{l}) \sin(\lambda \varphi) \frac{\partial}{\partial {r}} + \frac{1}{r} \sin(\frac{\lambda t}{l}) \cos(\lambda \varphi) \frac{\partial}{\partial {\varphi}}\Big],\nonumber\\
K_{3}  &=& \sqrt{\lambda^2 {l^2}+{r^2}}~~\Big[-\frac{r l}{\lambda^2 {l^2} +r^2}~\sin(\frac{\lambda t}{l}) \sin(\lambda \varphi)~\frac{\partial}{\partial {t}} \nonumber\\
&& +\lambda \cos(\frac{\lambda t}{l}) \sin(\lambda \varphi) \frac{\partial}{\partial {r}} + \frac{1}{r} \cos(\frac{\lambda t}{l}) \cos(\lambda \varphi) \frac{\partial}{\partial {\varphi}}\Big],\nonumber\\
K_{4}  &=& \sqrt{\lambda^2 {l^2}+{r^2}}~~\Big[-\frac{r l}{\lambda^2 {l^2} +r^2}~\cos(\frac{\lambda t}{l}) \cos(\lambda \varphi)~\frac{\partial}{\partial {t}} \nonumber\\
&& -\lambda \sin(\frac{\lambda t}{l}) \cos(\lambda \varphi) \frac{\partial}{\partial {r}} + \frac{1}{r} \sin(\frac{\lambda t}{l}) \sin(\lambda \varphi) \frac{\partial}{\partial {\varphi}}\Big],\nonumber\\
K_{5}  &=& \sqrt{\lambda^2 {l^2}+{r^2}}~~\Big[\frac{r l}{\lambda^2 {l^2} +r^2}~\sin(\frac{\lambda t}{l}) \cos(\lambda \varphi)~\frac{\partial}{\partial {t}} \nonumber\\
&& -\lambda \cos(\frac{\lambda t}{l}) \cos(\lambda \varphi) \frac{\partial}{\partial {r}} + \frac{1}{r} \cos(\frac{\lambda t}{l}) \sin(\lambda \varphi) \frac{\partial}{\partial {\varphi}}\Big],\nonumber\\
K_{6}  &=& -l^2 \frac{\partial}{\partial {t}}.\label{eq15}
\end{eqnarray}
Fortunately, for integer values of $\lambda$, the resulting Killing vectors are single-valued in the angular direction.
Now, we apply the Killing vectors (\ref{eq15}) to construct an appropriate vector field by using  formula (\ref{eq13}).
As we will see, our solutions are classified into two special cases. In both cases of solutions, the anti-symmetry
tensor field $B$ is considered to be $B=-{r^2}/{l}~ dt \wedge d\varphi$. Then,
one can employ formula (\ref{eq4}) to calculate the field strength corresponding to this $B$-field, giving us
\begin{equation}\label{eq17}
H=-\frac{2r}{l}~ dr \wedge dt \wedge d\varphi.
\end{equation}
Here, the forms of our solutions including the metric (\ref{eq14}) and field strength (\ref{eq17}) are given by the following two cases I and II:\\
$\bullet$~Case I: In this case, the GSE (\ref{eq1})-(\ref{eq3}) together with (\ref{eq5})-(\ref{eq8}) are fulfilled with the metric (\ref{eq14}) and field strength (\ref{eq17}) if the vector field $I$, one-form $Z$ and cosmological constant $\Lambda$ can now be expressed in the following forms
\begin{equation}\label{eq18}
I=-\alpha_{6} l^2~\frac{\partial}{\partial {t}},~~~~~Z=\alpha_{6} l r^2~ d\varphi,~~~~\Lambda =-\frac{1}{l^2} +\lambda^2  {\alpha_{6}}^2 l^4.
\end{equation}
Using these, the dilaton field is constant, and the components of $X_\mu$ read off
\begin{equation}\label{eq19}
X_t=\alpha_{6} l^2 (\lambda^2 +\frac{r^2}{l^2}),~~~~~X_\varphi=\alpha_{6} l r^2,~~~~~~X_r=0.
\end{equation}
$\bullet \bullet$~Case II: In this case of solutions, the vector field $I$ is a rotational symmetry,
which together with one-form $Z$ and  $\Lambda$ are given as follows:
\begin{equation}\label{eq20}
I=\alpha_{1}~\frac{\partial}{\partial {\varphi}},~~~~Z=\alpha_{1} l (\lambda^2 +\frac{r^2}{l^2})  dt,~~~
\Lambda =-\frac{1}{l^2} +\lambda^2  {\alpha_{1}}^2 l^2.~
\end{equation}
Then, one finds that
\begin{equation}\label{eq21}
X_t=\alpha_{1} l (\lambda^2 +\frac{r^2}{l^2}),~~~~~X_\varphi=\alpha_{1} r^2,~~~~~~X_r=0.
\end{equation}
The corresponding dilaton field to this case of the solutions is time-dependent, $\Phi=c_{_0} +\lambda^2  {\alpha_{1}} l t$.

Let us now discuss the solutions for all positive and negative values of $M$, when $J$ is zero.
The BTZ metric with $J=0, M \neq 0$ is given by
\begin{equation}\label{eq22}
ds^2=(M-\frac{r^2}{l^2})~dt^2+(\frac{r^2}{l^2}-M )^{-1}~dr^2+r^2~d\varphi^{2}.
\end{equation}
In this case, the solutions are similar to those of the case $J=0, M =-\lambda^2$, so that
one should use $M$ instead of $-\lambda^2$ in the solutions (\ref{eq18}) and (\ref{eq20}). Then,
the solutions take the forms
\begin{equation}\label{eq23}
I=-\alpha_{6} l^2~\frac{\partial}{\partial {t}},~~~~~Z=\alpha_{6} l r^2~ d\varphi,~~~~\Lambda =-\frac{1}{l^2} -M {\alpha_{6}}^2 l^4,
\end{equation}
and
\begin{equation}\label{eq24}
I=\alpha_{1}~\frac{\partial}{\partial {\varphi}},~~~~Z=\alpha_{1} l (\frac{r^2}{l^2} -M)  dt,~~~
\Lambda =-\frac{1}{l^2} -M  {\alpha_{1}}^2 l^2,~
\end{equation}
and in the same way for the components of $X_\mu$.

\subsection{Solution with J=0, M=0 }

The BTZ metric with $J=0,~M=0$ is simply found by using the formula (\ref{eq10}), giving us
\begin{equation}\label{eq25}
ds^2=-\frac{r^2}{l^2}~dt^2+\frac{l^2}{r^2}~dr^2+r^2~d\phi^{2}.
\end{equation}
In order to construct the vector field $I$, one may determine the Killing vectors corresponding to the metric
(\ref{eq25}) similar to what was done in (\ref{eq15}). Analogously,
only non-zero component of the tensor field $B$ is considered to be $B_{\varphi t} ={r^2}/{l}$ for which
the field strength is calculated to be as in (\ref{eq17}).
Now, we solve the GSE with the metric (\ref{eq25}), field strength (\ref{eq17}) and a constant dilaton field.
The equations are then satisfied if the vector field $I$, one-form $Z$ and $ \Lambda$ 
have the following forms\footnote{Note that our solution including the metric (\ref{eq25}), $B_{\varphi t} ={r^2}/{l}$ and relations given by
equation (\ref{eq26}) together with a constant dilaton field may be related to
a GSE solution found in equation (4.12) of Ref. \cite{Sakatani}. There, it has been obtained
a 10-dimensional solution for the GSE (with a non-zero Killing vector $I$) whose metric is locally $AdS_3 \times S^3 \times T^4$.
If we look at the $AdS_3$ part of solution (4.12) of \cite{Sakatani},
it may be locally the same as our BTZ solution.}

\begin{eqnarray}
 I&=&-\alpha_{6} l^2~\frac{\partial}{\partial {t}} + \alpha_{3}~\frac{\partial}{\partial {\varphi}},\nonumber\\
 Z&=&\frac{\alpha_{3} r^2}{l}  dt + \alpha_{6}  l r^2 d \varphi,\nonumber\\
 \Lambda &=& -\frac{1}{l^2}.\label{eq26}
\end{eqnarray}
Using the first two relations and also the $B$-field $B_{\varphi t} ={r^2}/{l}$, one then finds that the components of $X_\mu$ are
\begin{eqnarray}
X_t&=& (\alpha_{3} +\alpha_{6} l) ~\frac{r^2}{l},\nonumber\\
X_\varphi&=& (\alpha_{3} +\alpha_{6} l) ~r^2,\nonumber\\
X_r &=& 0. \label{eq27}
\end{eqnarray}

At the end of this section let us compare the solutions with J=0, M=0  and  J=0, M$\neq $0.
As showed in the above, for the case of J=0, M=0 we obtained the vector $I$ as a linear combination of the directions of the time translation and rotational symmetries, while this is not true for the case of J=0, M$\neq $0. In the case  M$\neq $0 of the solutions, if one uses a linear combination of
(\ref{eq23}) and (\ref{eq24}) for the vector $I$ and one-form $Z$, then, equation (\ref{eq8}) becomes $\alpha_{1} \alpha_{6} l^3 M =0$. This means that one of the constants $\alpha_{1}$ or $\alpha_{6}$ should be zero. Therefore, unlike the case of J=0, M=0,
a linear combination of (\ref{eq23}) and (\ref{eq24})
cannot be a solution for the GSE.

\section{\label{V} Abelian target space dual of the BTZ as a solution of the GSE}

One of the interesting issues in the context of GSE is that to investigate the solutions of the GSE
under the T-duality. The T-duality symmetry is one of the most interesting properties of string theory connecting seemingly different backgrounds in which
the strings propagate.
We say that the duality is Abelian if it is constructed on an Abelian isometry group \cite{Busher}.
By making use of the Buscher's duality transformations \cite{Busher}, it was shown that
the BTZ black hole solution discussed at the end of section \ref{III} is, under the Abelian T-duality, equivalent to
the charged black string solution discussed in Ref. \cite{Horowitz1}. There, Horne and Horowitz found
a family of solutions to low energy string theory describing charged black strings in three dimensions. This family of
solutions is given by
\begin{eqnarray}
{d {\tilde s}}^2 &=&-\big(1-\frac{{\bf M}}{\hat{r}}\big) d \hat{t}^2+ \Big(1-\frac{{{\bf Q}}^2}{{{M}} \hat{r}}\Big) d \hat{x}^2\nonumber\\
&&~~~~~~~~~~+\big(1-\frac{{\bf M}}{\hat{r}}\big)^{^{-1}}  \Big(1-\frac{{{\bf Q}}^2}{{{\bf M}} \hat{r}}\Big)^{^{-1}}~ \frac{l^2 d \hat{r}^2}{4 \hat{r}^2},\label{eq29}\\
{\tilde H} &=& \frac{{\bf Q}}{\hat{r^2}}~ d \hat{r} \wedge  d {\hat{t}} \wedge d \hat{x},\label{eq30}\\
{\tilde \Phi} & = & -\frac{1}{2} \ln (l \hat{r}),\label{eq31}
\end{eqnarray}
where ${\bf M} = {r_+^2}/l$ and ${\bf Q} = {J}/2$ are the respective the mass and charge of the black string.
It can be easily shown that the above metric admits two Killing vectors $\partial/\partial \hat{t}$ and $\partial/\partial \hat{x}$.
Here, we shall show that the metric (\ref{eq29}) and field strength (\ref{eq30}) can satisfy the GSE, if the
vector field I, one-form Z, dilaton field and cosmological constant  express as follows:
\begin{eqnarray}
 I&=&\alpha_{1} ~\frac{\partial}{\partial \hat{t}},\nonumber\\
 Z&=&-\frac{1}{2 \hat{r}} d \hat{r} +\alpha_{1} (\frac{{\bf M}}{{\bf Q}}-\frac{{\bf Q}}{\hat{r}}) d \hat{x},\nonumber\\
 { \Phi} &=& c_{_0} -\frac{1}{2} \ln \hat{r} +\alpha_{1} \frac{{\bf M}}{{\bf Q}} \hat{x},\nonumber\\
\Lambda & = & \frac{2}{l^2} + \frac{2 {\alpha_{1}}^2 ({\bf M}^2 -{\bf Q}^2)}{{\bf Q}^2}.\label{eq32}
\end{eqnarray}
In addition to this solution, one can see that the GSE (\ref{eq1})-(\ref{eq3}) together with (\ref{eq5})-(\ref{eq8}) are
fulfilled with the metric (\ref{eq29}) and field strength (\ref{eq30}) if the vector field $I$, one-form $Z$, dilaton field and cosmological constant can now be expressed in the following forms
\begin{eqnarray}
 I&=&\alpha_{2} ~\frac{\partial}{\partial \hat{x}},\nonumber\\
 Z&=&-\frac{1}{2 \hat{r}} d \hat{r} +\alpha_{2} {\bf Q} (\frac{1}{\hat{r}} -\frac{1}{{\bf M}}) d \hat{t},\nonumber\\
 {\Phi} &=& c_{_0} -\frac{1}{2} \ln \hat{r} -\alpha_{2} \frac{{\bf Q}}{{\bf M}} \hat{t},\nonumber\\
\Lambda & = & \frac{2}{l^2} + \frac{2 {\alpha_{2}}^2 ({\bf M}^2 -{\bf Q}^2)}{{\bf M}^2}.\label{eq33}
\end{eqnarray}
\section{\label{VI} Conclusion}
In this study, we have obtained some new solutions for the GSE in three dimensions.
Our solutions include the special cases $J= 0, M= 0$ and $J= 0, M \neq 0$ of the BTZ metric together with the
field strength $H=-({2r}/{l}) dr \wedge dt \wedge d\varphi$.
In each case we have calculated the the vector field $I$, one-form $Z$ and the cosmological constant $ \Lambda$.
To determine the vector field $I$ we have used the Killing vectors corresponding to the BTZ metrics.
By comparing the solutions with $J= 0, M= 0$ and $J= 0, M \neq 0$, we concluded that unlike the case of $J=0, M=0$,
a linear combination of (\ref{eq23}) and (\ref{eq24}) cannot be a solution for the GSE.
However, we have derived a family of solutions as presented in relations (\ref{eq23}), (\ref{eq24}) and (\ref{eq26}).
Notably, our results confirm that these solutions are single-valued in the angular direction.
On the other hand, we have checked that the cases $J\neq 0, M=0$ and $J\neq 0, M\neq 0$ for the BTZ metric do not satisfy the GSE.

While we considered the BTZ metrics, one can easily repeat for other geometric metrics such as Thurston geometries.
As mentioned in section \ref{III}, the BTZ metric with $J = 0$
and $M = -1$ represents ordinary anti-de Sitter
space ($AdS_3$), which is nothing but one of Lorentzian Thurston geometries. Therefore, we have investigated that the $AdS_3$ space can be considered as a solution for the GSE.
Finally, as an interesting result, we have shown that the charged black string solution
(\ref{eq29}), which is Abelian T-dual to the the BTZ black hole solution, can be also a solution for the GSE.
In fact, this result helps to answer the question whether the solutions of the GSE are, under the Abelian T-duality, preserved.
It will be an interesting future direction to be addressed.

\begin{acknowledgements}
This work has been supported by the research vice chanceller of Azarbaijan Shahid Madani University under research fund No. $1402/231$.
\end{acknowledgements}



\begin{thebibliography}{9}
		
\bibitem{r1}
G. Arutyunov, S. Frolov, B. Hoare, R. Roiban, A. A. Tseytlin, Nucl. Phys. B \textbf{903}, 262 (2016)

\bibitem{r2}
F. Delduc, M. Magro, B. Vicedo,  Phys. Rev. Lett.  \textbf{112}, 051601 (2014)
		
\bibitem{r3}
F. Delduc, M. Magro, B. Vicedo,  J. High Energy Phys. \textbf{10}, 132 (2014)
		
\bibitem{r4}
I. Kawaguchi, T. Matsumoto, K. Yoshida, J. High Energy Phys.  \textbf{04}, 153 (2014)

		
\bibitem{r5}
R. Borsato, L. Wulff, J. High Energy Phys. \textbf{10}, 045 (2016)

\bibitem{Sheikh-Jabbari1}	
T. Araujo, I. Bakhmatov, E. O Colgain, J. Sakamoto, M. M. Sheikh-Jabbari, K. Yoshida,
Phys. Rev. D {\bf 95}, 105006 (2017)


\bibitem{Sheikh-Jabbari2}
T. Araujo, I. Bakhmatov, E. O Colgain, J. Sakamoto, M. M. Sheikh-Jabbari, K. Yoshida,
 J. Phys. A {\bf 51}, 235401  (2018)

\bibitem{Sheikh-Jabbari3}
T. Araujo, E. O Colgain, J. Sakamoto, M. M. Sheikh-Jabbari, K. Yoshida,
Eur. Phys. J. C \textbf{77}, 739 (2017)


\bibitem{Sheikh-Jabbari4}
 I. Bakhmatov, O. Kelekci, E. O Colgain, M. M. Sheikh-Jabbari,
 Phys. Rev. D {\bf 98}, 021901 (2018)

	
\bibitem{r6}
D. Orlando, S. Reffert, J. I. Sakamoto, K. Yoshida, J. Phys. A \textbf{49}, 445403 (2016)
		

\bibitem{r7}
B. Hoare, A. A. Tseytlin, J. Phys. A \textbf{49}, 494001 (2016)
		

\bibitem{r8}
M. Hong, Y. Kim, E. O. Colgain, Eur. Phys. J. C \textbf{78}, 1025 (2018)

\bibitem{r9}
R. Borsato, L. Wulff, J. High Energy Phys. \textbf{08}, 027 (2018)


\bibitem{r11}
B. Hoare, A. A. Tseytlin,  J. High Energy Phys. \textbf{10}, 060 (2015)

\bibitem{r10}
L. Wulff, A. A. Tseytlin,  J. High Energy Phys. \textbf{06}, 174 (2016)
		
\bibitem{yuho1}
J. Sakamoto, Y. Sakatani, K. Yoshida,
PTEP {\bf 053B07} (2017)

\bibitem{yuho1.1}
Y. Sakatani, S. Uehara, K. Yoshida,
J. High Energy Phys. \textbf{04}, 123 (2017)


\bibitem{yuho2}
J. J. Fernandez-Melgarejo, J. Sakamoto, Y. Sakatani, K. Yoshida,
 Phys. Rev. Lett. {\bf 122}, 111602 (2019)

\bibitem{Horowitz1}
J. Horne and G. Horowitz, Nucl. Phys. B {\bf 368}, 444 (1992)

\bibitem{Horowitz}
G. Horowitz, D. Welch, Phys. Rev. Lett. \textbf{71}, 328 (1993)

\bibitem{r12}
M. Banados, C. Teitelboim, J. Zanelli, Phys. Rev. Lett. \textbf{69}, 1849 (1992)

\bibitem{Sakatani}
Y. Sakatani, Prog. Theor. Exp. Phys. {\bf 073}B04 (2019)

\bibitem{Busher}
T. Buscher,  Phys. Lett.  B {\bf 194}, 59 (1987);  Phys. Lett.  B {\bf 201}, 466 (1988).


\end{thebibliography}
\end{document}